\documentclass[prd,preprint
               showpacs,                                       
               preprintnumbers,aps,epsfig,
               nofootinbib]{revtex4}
\usepackage{graphicx}
\usepackage{amssymb,latexsym}

\newcommand{\order}{{\cal O}}
\newcommand{\beq}{\begin{equation}}
\newcommand{\eeq}{\end{equation}}
\newcommand{\beqa}{\begin{eqnarray}}
\newcommand{\eeqa}{\end{eqnarray}}

\newcommand{\lmk}{\left(}
\newcommand{\rmk}{\right)}
\newcommand{\lkk}{\left[}
\newcommand{\rkk}{\right]}

\newcommand{\GeV}{~\mbox{\rm GeV}}

\begin{document}

\preprint{RESCEU-2/08,UTAP-594}

\date{\today}

\title{%
 Space laser interferometers can determine\\ 
the thermal history of the early Universe
}
\author{%
 Kazunori Nakayama$^1$, Shun Saito$^2$, Yudai Suwa$^2$, 
and Jun'ichi Yokoyama$^3$
}
\address{%
$^1$Institute for Cosmic Ray Research, The University of Tokyo,
Kashiwa, Chiba 277-8582, Japan\\
$^2$Department of Physics, Graduate School of Science,  
The University of Tokyo, 
Tokyo 113-0033, Japan\\
$^3$Research Center for the Early Universe (RESCEU),
Graduate School of Science, 
The University of Tokyo,
Tokyo 113-0033, Japan\\
}

\begin{abstract}
It is shown that space-based gravitational wave detectors such 
as DECIGO and/or Big Bang Observer (BBO) will provide us with 
invaluable information on the cosmic thermal history after inflation 
and they will be able to determine the reheat temperature $T_R$
provided that it lies in the range preferred by the cosmological
gravitino problem, $T_R\sim 10^{5-9}$ GeV.  
Therefore it is strongly desired that they will be put into practice 
as soon as possible. 
\end{abstract}


\maketitle


Although we can probe physics during inflation in the 
early Universe \cite{lindebook} 
rather precisely now using observations 
of the anisotropy in the cosmic microwave 
background (CMB) radiation \cite{WMAP3COSMO}, 
cosmic evolution from the end of inflation to the beginning of
the primordial big-bang nucleosynthesis (BBN)  is still in the dark age, 
when many important phenomena  such as (p)reheating,
baryogenesis, generation and freeze-out of cold dark matter particles, 
etc.\ have taken place. It is desirable to clarify 
the cosmic thermal history in this regime observationally. 

Here we argue that future space-based gravitational wave detectors 
such as DECIGO \cite{decigo} or the Big Bang Observer (BBO), 
are very useful for this purpose and may  be able to determine 
the reheat temperature after inflation by observing stochastic
gravitational radiation background generated during inflation in 
the frequency range 
around $0.1-10$ Hz where foregrounds from astrophysical objects
are separable \cite{Farmer}
\footnote{
  Recently, Population III stars are proposed as a dominant component
  around the deci-Heltz band as a result of the gravitational
  radiation associated with neutrino emissions \cite{Buonanno:2004tp}.
  However, the amplitude is very uncertain due to the lack of
  understating on early star formation history.  This signal would be
  separable if we adopt reliable abundance of Population III stars and
  take the duty cycle and their angular distribution into account. }.

We introduce tensor perturbations, $h_{ij}$, around a spatially flat
Robertson-Walker metric as 
\beq
ds^2=-dt^2+a^2(t)\left( {\delta _{ij}+2h_{ij}} \right)dx^idx^j,
\eeq
with $a(t)$ being the scale factor.
Decomposing the tensor metric perturbation to  Fourier modes as
\beq
h_{ij}=\sqrt {8\pi G}\sum\limits_{A=+,\times } {\int {{{d^3k} \over
{(2\pi )^{{3 \mathord{\left/ {\vphantom {3 2}}
\right. \kern-\nulldelimiterspace} 2}}}}}}\varphi _k^A(t)
e^{i{\bf k}\cdot{\bf x}}e_{ij}^A,
\eeq
we find that the two independent degrees of freedom $\varphi^{A}$ behave as
two massless minimally coupled scalar fields, where $e_{ij}^A$
represents polarization tensor with 
$e_{ij}^Ae_{}^{ijA'}=\delta ^{AA'}$ for $A,~A'=+,\times$.
Applying quantum field theory of a massless minimally coupled field in
de Sitter spacetime, we find that the Fourier modes are characterized by the
following vacuum correlation,
\beq
\left\langle {\varphi _k^A(t)\varphi _{k'}^{A'}(t)} \right\rangle ={{H^2}
\over {2k^3}}\delta ^3\left( {k-k'} \right)\delta ^{AA'},
\eeq
so that the amplitude per logarithmic frequency interval is given by
\beqa
h_{F} ^2 (f ) &\equiv& 2\left\langle {h_{ij} h^{ij} (f )} 
\right\rangle  = 
4 \times 8\pi G\left( {\frac{{H(\phi) }}{{2\pi }}}\right)^2 \\
 &=& \frac{V[\phi(f)]}{3\pi^2 M_G^4}
\equiv \frac{1}{2}\Delta_h^2(f), \nonumber
\eeqa
in the long wavelength regime.  Here
 $M_G=(8\pi G)^{-1/2}$ is the reduced Planck mass and
$V[\phi(f)]$ is the value of potential energy density of
the inflaton when 
the scale corresponding to the frequency $f$ today left the
Hubble radius during inflation.  $\Delta_h$ is an 
expression for the amplitude of tensor perturbation used frequently
in the analysis of cosmological observations.

The above expression gives the initial condition
 to the solution of each Fourier 
mode in the post inflationary universe which behaves as
\beq
 h(f,a)\propto
a(t)^{\frac{1-3p}{2p}}J_{\frac{3p-1}{2(1-p)}}\lmk\frac{p}{1-p}
\frac{k}{a(t)H(t)}\rmk,~~~k=2\pi fa_0, \label{7}
\eeq
in a power-law background
 $a(t)\propto t^p$ with $p<1$.  Here $J_n(x)$ is a Bessel function and
 $a_0$ denotes the current scale factor.

Thus the amplitude of 
gravitational
wave takes a constant value,
$h(f,a)=h_F(f)$, until its wave length falls shorter than
 the Hubble radius $H^{-1}$ at 
$a=2\pi fa_0/H \equiv a_{in}(f)$.
From the asymptotic expansion of (\ref{7}), one can see that
the energy density 
stored in the tensor perturbation starts to decrease just as 
radiation once the wavelength falls to a sub-horizon scale.  
Therefore in this regime its relative energy density to the 
background density remains constant during radiation 
domination, while it decreases when 
the Universe is dominated by other form of energy such as 
nonrelativistic matter or coherent field oscillation.
If such a stage lasts long, 
energy density in sub-horizon tensor perturbations tends to 
be suppressed and the resultant spectrum is modified from
a nearly scale-invariant one.
The present density parameter of the gravitational radiation
per logarithmic frequency interval is described as 
\beq
 \Omega_{GW}(f,t_0)
= \frac{(2\pi f)^2}{12H_0^2}\Delta_h^2(f)\lmk\frac{a_{in}(f)}{a_0}\rmk^2. 
\label{infnorm}
\eeq 
It behaves as $\Omega_{GW}(f,t_0)\propto f^{-2} (f^0)$
for the mode which enters the horizon 
in the matter (radiation) dominated regime.
Thus the stochastic gravitational wave background not 
only carries information on the inflationary regime, during 
which they are generated, but also serves as a probe of the 
equation of state in the early universe \cite{seto}
\if
This double role played by 
primordial gravitational wave background in particle cosmology
is similar to that played by high-redshift quasars in 
observational cosmology, which not only reflects the 
properties of high-redshift universe at their location, 
but also carry line-of-sight information in their absorption 
spectra.  In this sense, just as high-redshift quasars are 
regarded as lighthouses of the distant universe, we may 
regard the inflation-produced gravitational wave background 
as a lighthouse that can shed light on the early Universe
which serves as a laboratory for high energy 
physics
\fi  
(see also \cite{others} for other applications of stochastic
gravitational wave background ). 

Besides direct observation by space laser interferometers,
cosmological 
stochastic gravitational wave background can be detected
indirectly through CMB. 
By measuring its B-mode polarization, we can probe the amplitude
of tensor perturbation from inflation.

In a simple single-field slow-roll inflation model with a 
potential $V[\phi]$, observable quantities such as the 
amplitude of curvature perturbation, $\Delta_{\cal R}$,
its spectral index, $n_s$, and running, $dn_s/d\ln k$,
are described by the slow-roll parameters,
\[
  \epsilon=\frac{M_G^2}{2}\lmk\frac{V'[\phi]}{V[\phi]}\rmk^2,~
\eta=M_G^2\frac{V''[\phi]}{V[\phi]},~
\xi=M_G^4\frac{V'[\phi]V'''[\phi]}{V^2[\phi]},
\]
as
\[
 \Delta_{\cal R}^2=\frac{V[\phi]}{24\pi^2M_G^4\epsilon},~
 n_s=1-6\epsilon+2\eta,~
\frac{dn_s}{d\ln k}=16\epsilon\eta-24\epsilon^2-2\xi,
\]
where each quantity should be evaluated at the time of
horizon crossing during inflation.

WMAP has measured both $\Delta_{\cal R}$ and $n_s$ with an unprecedented
accuracy.  However, since $\Delta_{\cal R}$ depends not only on
$V[\phi]$ but also on the slow-roll parameter $\epsilon$, we have
not been able to fix the energy scale of inflation yet.
Detection of tensor perturbation is essentially important to 
determine the energy scale of inflation.
\if
The importance of the observation of tensor mode to fix the 
tensor-to-scalar ratio $r\equiv \Delta_{h}^2/\Delta_{\cal R}^2$
lies indeed in the fact that we can determine the energy scale
or the epoch of inflation.
\fi

Theoretically, different inflation models predict different 
energy scale of accelerated expansion.  Among them, chaotic inflation
model \cite{chaoinf}
with a massive scalar potential, $V[\phi]=m^2\phi^2/2$,
which is attractive both from naturalness of initial condition
and phenomenological point of view \cite{Murayama:1992ua}, 
we find a relatively large
amplitude of tensor perturbation: 
$r\equiv \Delta^{2}_{h}/\Delta^{2}_{\mathcal R}(k_{0})\cong 0.16$, 
where pivot scale, $k_{0}$, corresponds to the CMB scale. 
Other models
such as small field models may predict much smaller value of $r$
including a vanishingly small one, which would make us desperate
with regards to the detection of B-mode polarization.

Recently, however, it has been claimed that, if we use the observed
value of the scalar spectral index $n_s=0.961\pm 0.017$ by WMAP
as a constraint, even small-field models would predict $r>10^{-3}$
\cite{Pagano}.  That is, it would be unusual to have 
$\eta \gg \epsilon$ because it would mean the inflaton is near
an inflection point when the CMB scale left the Hubble radius
and it would be very difficult
to achieve the right number of $e$-folds thereafter without
severe fine tuning.  Then the observed
deviation of the spectral index from unity, $n_s\simeq 0.95-0.99$,
implies $\epsilon=\order (0.01)$ or $r\sim 0.1$.  Indeed authors 
of \cite{Pagano}
  have considered a number of potentials with different parameters
and calculated $n_s$ and $r$ under the condition that the
duration of inflation takes a proper value.  As a result they
find that the observed value of $n_s$ implies  $r> 0.003$.

The same issue has been studied by Boyle, Steinhardt, and Turok 
\cite{Boyle:2005ug} in somewhat
more model-independent manner.  They count the number of zeros of
the slow roll parameters $\epsilon$ and $\eta$ in the last 60
 $e$-folds of inflation as a conservative measure of how many
 derivatives of them must be fine tuned to achieve a given set of
$(n_s,r)$.  They find inflation models with no fine tuning
give $n_s < 0.98$ and $r>10^{-2}$ and that models predicting
$r< 10^{-3}$ require nine or more extra degrees of fine tuning.

If $r$ is indeed larger than, say, 0.003, we will be able to detect
B-mode polarization by ongoing and planned projects and its implication
is profound.  Because we already know that $\Delta_{\cal R}^2\simeq
2.0\times 10^{-9}$ on CMB scale \cite{WMAP3COSMO}, by measuring $r$
we can fix the energy scale of inflation as
\[
 V[\phi]=\lmk 3.2\times 10^{16}\rm GeV\rmk^4r=
\lmk 7.5\times 10^{15}\rm GeV\rmk^4\lmk\frac{r}{0.003}\rmk .
\]

Once we succeed in measuring $r$ by B-mode polarization, we will be
even more confident in simple slow-roll single-field inflation, on which
the above arguments \cite{Pagano,Boyle:2005ug} are  based,
and we can predict the amplitude of gravitational waves on frequencies
 accessible by the other means of observation, namely, 
a space laser interferometer such as DECIGO \cite{decigo} and BBO.
Extrapolating the amplitude of tensor perturbation on the CMB
scale with the wavenumber $k_0=0.002{\rm Mpc^{-1}}$ corresponding
to the frequency $f_h=3\times 10^{-18}$Hz today, to higher 
frequency using the slow-roll parameters measured at $k_0$ we find
\begin{widetext}
\beq
  \Delta_h^2(f)=\Delta_h^2(f_h)\lkk 1-2\epsilon\ln\frac{f}{f_h}
+2\epsilon(\eta-\epsilon)\lmk\ln\frac{f}{f_h}\rmk^2 +\frac{1}{3}\epsilon
(-12\epsilon^2+16\epsilon \eta-4\eta^2-2\xi)\lmk\ln\frac{f}{f_h}\rmk^3 \rkk.
\eeq
\end{widetext}

Among the planned space laser interferometers, LISA \cite{LISA}
is most sensitive to frequency around $10^{-3}$ Hz where stochastic 
gravitational wave background is dominated by astronomical sources 
such as white dwarf binaries and it seems difficult to detect 
inflationary gravitational wave background even if $r$ is maximal.
On the other hand, DECIGO or BBO targets frequency window 
around $f=0.1-1$ Hz where contamination of astrophysical 
foregrounds is separable, so they are ideal to probe inflation.
In the standard cosmology the gravitational wave with its current 
frequency 0.1 Hz reentered the Hubble radius in the radiation dominated
regime when the cosmic temperature
was $T=4\times 10^6$ GeV assuming that there were no significant entropy 
production after that.
  
Interestingly, this temperature is in the expected range of the
upper bound on the reheat temperature, $T_R$, after inflation imposed by 
the decay of unstable gravitinos which are produced by thermal
scattering in the reheating stage. 
Specifically, in order to ensure
successful BBN in the presence of the hadronic decay of gravitinos,
the reheat temperature should satisfy $T_R \lesssim 10^{6-8}$GeV 
depending on the gravitino mass and the hadronic branching ratio
\cite{Kawasaki:2004qu}.  In case gravitino is the lightest
supersymmetric particle and hence stable due to R-parity conservation,
we also find a constraint on the reheat temperature so that
it does not overclose the universe,
\beq
 T_R < 7\times 10^6\lmk\frac{m_{\tilde g}}{1~{\rm TeV}}\rmk^{-2}
 \lmk\frac{m_{3/2}}{1 \GeV}\rmk \GeV,
\eeq
for the gravitino mass $m_{3/2}=10^{-4}-10$ GeV  \cite{Moroi:1993mb},
where $m_{\tilde g}$ denotes the gluino mass.

If the reheat temperature indeed satisfies the above constraints,
it is likely that the universe is still dominated by coherent inflaton
field oscillation when the frequency range probed by DECIGO/BBO
reentered the Hubble radius.  Since the cosmic expansion law is the same
as in the matter-dominated era as field oscillation is driven
by a mass term, the resultant spectrum of $\Omega_{GW}(f,t_0)$ acquires
a modulation proportional to $f^{-2}$ for those frequency bands 
entering the horizon in this regime.

So far we have implicitly assumed that there is no significant
entropy production after the completion of reheating after inflation.
If nonnegligible amount of entropy is produced in the lower temperature
regime, from the decay of a long-lived scalar field other than the
inflaton for example, not only the spectrum of $\Omega_{GW}(f,t_0)$
but also constraint on the reheat temperature imposed by the thermal
gravitino problem are modified.  
Defining the entropy increase factor
$F$ by the ratio of entropy in a fixed comoving volume before and after
the entropy production \cite{seto}, 
we find that the overall amplitude
of the high-frequency part of $\Omega_{GW}(f,t_0)$, the
frequency of gravitational radiation entering the horizon at the 
end of reheating, and the constraint on the reheat temperature by the
gravitino problem are multiplied by $F^{-4/3}$, $F^{-1/3}$, and $F$,
respectively.

Taking all these factors into account we can numerically calculate
the spectrum of stochastic gravitational wave background in the band
probed by DECIGO/BBO using future-observed values of the
tensor-to-scalar ratio and the slow-roll
parameters.  Figure 1 summarizes the result of numerical calculation
for the case $r=0.16$,  $\epsilon=\eta=0.01$, and $\xi=0$, which are
realized in chaotic inflation driven by a quadratic potential.

We find a nearly flat spectrum
\beqa
 \Omega_{GW}(f,t_0)\!\! &=& \!\!
2.8\times 10^{-16}F^{-4/3}g_{\ast 106.75}^{-1/3}
\lmk\frac{r}{0.1}\rmk 
\lkk \frac{\Delta_h(f)}{\Delta_h(f_h)}\rkk^2 \nonumber \\
&\equiv & \Omega_{GW}^{(L)}(f,t_0),~~
\eeqa 
in the low frequency band reentering the horizon in the radiation
dominated regime after reheating (and before possible entropy 
production),
\beq
 f \lesssim 0.042\lmk\frac{T_RF^{-1/3}}{10^7 \GeV}\rmk~{\rm Hz}\equiv f_-.
\eeq
Here $g_{\ast 106.75}$ denotes effective number of relativistic
degrees of freedom normalized by 106.75 
when the relevant mode reentered the Hubble radius.
For higher frequency modes with
\beq
 f \gtrsim 1.4\lmk\frac{T_RF^{-1/3}}{10^7 \GeV}\rmk~{\rm Hz}\equiv f_+,
\eeq
we obtain a nearly power-law spectrum
\beqa
  \Omega_{GW}(f,t_0)&=&\Omega_{GW}^{(L)}(f,t_0)\lmk\frac{f_T}{f}\rmk^2, \\
  f_T &\equiv& 0.31\lmk\frac{T_RF^{-1/3}}{10^7 \GeV}\rmk~{\rm Hz}.
\eeqa
\begin{figure}
\includegraphics[scale=0.5]{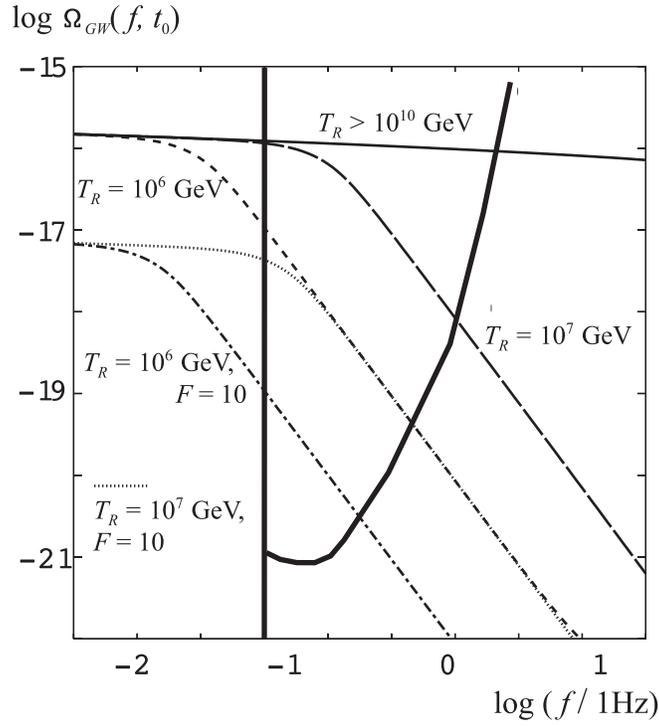}
\caption{Density parameter of gravitational radiation for different
reheat temperature $T_R$ and entropy increase factor $F$ in chaotic
inflation with a massive scalar field.  Curves without $F$
imply $F=1$.  The region above the thick curve with $f\gtrsim 0.1$Hz
can be observable by the ultimate DECIGO after its ten years'
operation. 
}
\label{fig:clbin}
\end{figure}
The cosmological information we can obtain by observing these
gravitational wave background depends on the shape of the 
spectrum observable at the DECIGO/BBO band, namely, 0.1Hz $\lesssim f
\lesssim$ 10 Hz, which can be classified to the following three 
cases {\bf (a)-(c)}.

{\bf  {Case (a)}} $f_- \gtrsim$ 10 Hz corresponding to 
$T_R \gtrsim 2.4\times 10^9 F^{1/3}\GeV$:
We can observe the nearly-flat region of the spectrum and determine
$F$ from the overall amplitude of $\Omega_{GW}$.  Since WMAP constrains
$r < 0.55$ \cite{WMAP3COSMO}, $F$ should satisfy 
$F \lesssim 3\times 10^4$.  
Thus for stable gravitino, the above lower bound on $T_R$ indicates
\beq
  m_{3/2} > 0.4 \lmk\frac{F}{3\times 10^4}\rmk^{-2/3}
 \lmk\frac{m_{\tilde{g}}}{1~{\rm TeV}}\rmk^{2}\GeV,
\eeq
and for unstable one, larger mass $(m_{3/2}\gtrsim 10$~TeV) is preferred.

{\bf  {Case (b)}} $f_- \lesssim 10$ Hz and $f_+ \gtrsim 0.1$ Hz 
corresponding to $7.1\times 10^5F^{1/3}\GeV < T_R < 2.4\times
10^9F^{1/3} \GeV$:
We can fix $F$ from the overall amplitude and $T_R$ from the shape
of the spectrum independently.  This is the ideal case that 
space laser interferometers can practically determine the 
entire thermal history of the universe between inflation and
BBN.

{\bf  {Case (c)}} $f_+ \lesssim 0.1$ Hz
 corresponding to $  T_R < 7.1\times
10^5F^{1/3}\GeV$:
We can observe only the power-law region of the spectrum
and measure only the ratio $T_R/F$.  This ratio, however, 
fixes the gravitino-to-entropy ratio uniquely apart from a
logarithmic correction.

In conclusion, space laser interferometers such as DECIGO and BBO
will bring about invaluable information on the delayed 
reheating stage required by the gravitino problem after inflation 
and will be able to determine the reheat temperature and/or 
the entropy increase factor.
Hence it is desired that they are put into practice as soon as possible.

\acknowledgements 
The authors are grateful to Naoki Seto, Fuminobu
Takahashi, Atsushi Taruya, and Asantha Cooray for useful
communications.  This work was partially supported by JSPS through
research fellowships (KN, YS) and Grant-in-Aid for Scientific Research
Nos.~16340076(JY) and 19340054(JY).

\end{document}